\documentclass[prl,twocolumn,floatfix,showpacs,superscriptaddress]{revtex4}
\usepackage{graphics}
\usepackage{epsfig}
\usepackage{times}

\begin{document}
\preprint{HEP/123-qed}
\title{Grand Canonical Finite Size Numerical Approaches : a Route to Measuring Bulk Properties under Applied Field}
\author{Chisa Hotta}
\affiliation{Department of Physics, Kyoto Sangyo University, 
Kyoto 603-8555, Japan}
\author{Naokazu Shibata}
\affiliation{Department of Physics, Tohoku University, Sendai 980-8578, Japan}

\date{\today}
\begin{abstract}
We exploit a prescription to observe directly the physical properties of 
the thermodynamic limit under continuously applied field in one-dimensional quantum finite lattice systems. 
By systematically scaling down the energy of the Hamiltonian of the open system from center 
toward both ends, one could adopt the edge sites with negligibly small energy scale as 
grand canonical small particle bath, and an equilibrium states with non-integer arbitrary conserved numbers, 
e.g., electron numbers or $s_z$, are realized in the main part of the system. 
This will enable the evaluation of response functions under continuously varying external field in 
{\it a small lattice without any fine tuning or scaling of parameters} while keeping the standard numerical accuracy. 
Demonstrations are given on quantum spin systems and on a Hubbard model by the density matrix renormalization group. 
\end{abstract}
\maketitle
\narrowtext 
Physical properties of condensed matters have always been clarified in experiments by measuring their responses to applied fields, 
e.g. phase transition as a divergence of susceptibility under gradually varying conjugate fields, 
charge gaps by pumping up the electrons with varying frequencies in spectroscopies. 
Such "measurement", when applied to theories, requires an extra prescription; 
setting the system size, $L$, and quantizing it with the virtual boundaries, which discitizes the energy levels by $\sim \hbar/L$. 
By taking $L\rightarrow\infty$, the observables are extrapolated to their bulk values. 
This size scaling was practically indispensable from the early milestone numerical calculation 
by Bonner and Fisher on the magnetic susceptibility of spin chains\cite{bonner64}. 
Ever since then, how to reach larger $L$ and to find an appropriate scaling function was the standard direction of pursuing bulk results.
However, even far developed numerics at present are still unable to clarify numerous quantum many body problems,
particularly in two dimensions, where the size scaling is extremely difficult. 
Thid Communication develops a first-step-prescription to overcome this fundamental quantum mechanical problem in theories. 
The highlight is that one could directly observe physical quantities mimicking their thermodynamic limit at small fixed $L$. 
The observables and quantum numbers are {\it continuous} functions of applied fields, 
which enables the determination of the response functions to arbitrary small variation of fields, 
e.g. the differential susceptibility. 
In our setup, the continuity of the observables is guaranteed by using the system edges as virtual "particle bath" 
which is connected to the main part of the system by the small fluctuations. 
The numerical accuracy of the observables are insensitive to the particle number given on the whole cluster, 
since the excess particles from the required bulk value is absolved by the edge "particle bath".  
\par
For simplicity, we confine ourselves to one-dimensional (1D) quantum many body lattice models, 
and test the applicability of our scheme by comparing its demonstration by density-matrix renormalization group(DMRG) \cite{white92,review} with the exact results. 
However, the present scheme could be extended to higher dimensional systems or applied to any other numerical methods. 
Let us start from the general Hamiltonian on a 1D lattice consisting of $L$ sites and with two open ends, 
$ {\mathcal H} = \sum_{i=1}^{L}  u(i) +\; \sum_{l} \sum_{i=1}^{L-l}  h_l(i), $
where $u(i)$ includes the on-site interaction and potential, 
and $h_l(i)$ is the interaction between $i$-th and $(i+l)$-th sites. 
The major point of our setup is to deform the Hamiltonian as, 
\begin{eqnarray}
{\mathcal H}_{\rm deform} &=& \sum_{i=1}^{L} f_0(i) u(i) 
+\; \sum_{l}\sum_{i=1}^{L-l} f_l(i) h_l(i). 
\label{deform}
\end{eqnarray}
 by externally given function, $f_l(i)$, which should smoothly vary from the maximum value near the center to zero at both ends, 
so as to gradually scale down the energy. 
\begin{figure}[tbp]
\begin{center}
\includegraphics[width=6.5cm]{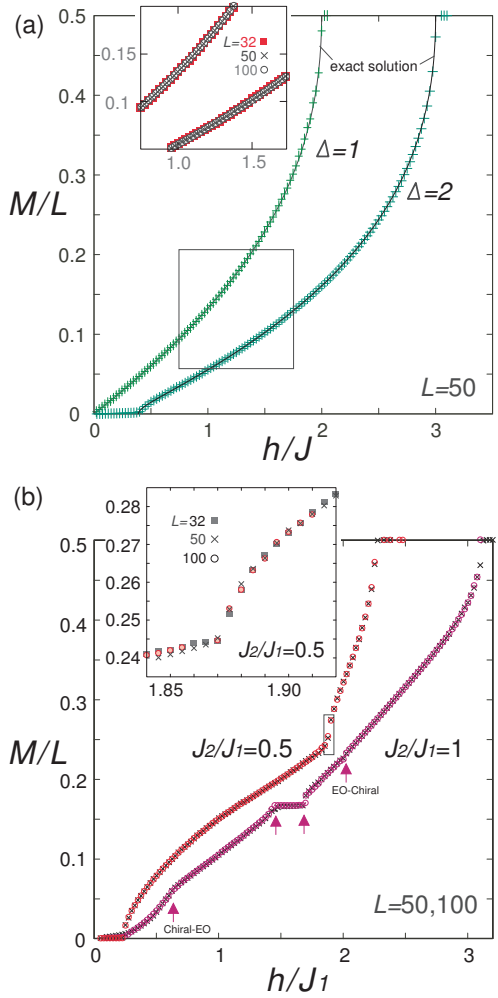}
\end{center}
\caption{(a) Magnetization curve of the $S=1/2$ Heisenberg ($\Delta=1$) and XXZ ($\Delta=2$) spin chain 
as a function of magnetic field $h$, with $J=1$. 
The results are obtained by the DMRG with $m \le 200$ on $\mathcal H_{\rm SSD}$ by our analyses. 
The solid line is the exact solution. 
(b) Magnetization curve of the $S=1/2$ $J_1$-$J_2$ Heisenberg model with $J_2/J_1=0.5$ and 1 obtained in the same manner as (a) with $m \le 300$. 
Arrows in the main panel indicate the phase transition (see. Ref.\onlinecite{okunishi08} for details.)
The inset shows the magnified curve near the cusp at $h\sim 1.8$ for $J_2/J_1$=0.5 for several system sizes. 
}
\label{f1}
\end{figure} 
%
The smooth boundary condition is first introduced to realize the flat translationally invariant wave function by getting rid of the boundary effect\cite{vekic93,vekic96}. 
Such a flat wave function is recently realized systematically by another function called sine-square deformation (SSD) with $f_l(i)= \sin^2\left(\frac{\pi (i+(l-1)/2)}{L}\right)$, which turned out to suppress the finite-size effects. 
Indeed, in the critical system, this SSD Hamiltonian, ${\mathcal H}_{\rm ssd}$, realizes a wave function of PBC \cite{katsura11,katsura11-2,hikihara11}, 
whose reason is partially clarified\cite{maruyama11,shibata11}. 
In our framework, we mainly adopt this SSD as a representative $f_l(i)$, since it is an established boundary and does not include adjustable parameters. 
However, notice that $f_l(i)$ is not limited to SSD, since the translational invariance/flatness of the wave function is not required in our scheme. 
{\it We confirmed that any function as far as it is convex downward at the edges, could be applied}(see the results in Fig.2(e)), 
We take full advantage regarding $f_l(i)$ that the energy and quantum fluctuation are all scaled down to nearly zero at the edges, 
and use these edges as buffers to absorb the deviation of energies and particle numbers from the thermodynamic value in the main part of the system. 
Using this setup, we develop an unprecedented method to calculate the static "bulk" responses to applied field. 
\par
In the following, the on-site field, $u(i)$ in Eq.(\ref{deform}), plays an important role; the Zeeman term,
$-h f_0(i) S^z_i$, where $h$ is the external magnetic field 
and $S^z_i$ is the $z$-component of the spin operator on site-$i$,
and the chemical potential term, $-\mu f_0(i) n_i$, for the electron system where $n_i$ is the electron number operator. 
In the SSD Hamiltonian, the fine tuning of $\mu$ was indispensable to recover the translational invariance 
in the electronic system off half-filling\cite{gendiar11}. 
In contrast, the present analyses no longer requires such fine tuning. 
\par
We start by showing the main results of our analyses in Fig. \ref{f1}(a), the magnetization curve of the $S=1/2$ Heisenberg spin chain and the XXZ spin chain, 
calculated by DMRG with $m \le 200$ where $m$ is the number of states kept per block. 
These models have $h_1(i)= J (S^x_iS^x_{i+1}+S^y_i S^y_{i+1}+ \Delta S^z_iS^z_{i+1})$, with $\Delta=1$ and $\Delta \ne 1$, respectively, 
where $S^\alpha$ is the $\alpha=x,y,z$ component of the spin operator. 
Both $\Delta=1$ and $2$ curves, calculated for $L=32,\;50$ and 100 coincide with the exact solutions\cite{exact} given in solid lines within the accuracy of ${\mathcal O}(10^{-5})$. 
The same calculation is also applied to the $J_1$-$J_2$ Heisenberg spin chain with $h_1(i)=J_1 S_i\cdot S_{i+1}$ and $h_2(i)= J_2 S_i\cdot S_{i+2}$. 
Figure~\ref{f1}(b) shows the magnetization curve for $J_2/J_1=0.5$ and 1. 
The former exhibits a characteristic cusp on a hillside \cite{okunishi97}, which is reproduced for several different choices of $L$ as shown in the inset. 
The latter curve at $J_2/J_1=1$ shows a clear 1/3 plateau and several anomalies indicating the quantum phase transitions. 
Here, we shall stress that the one above the plateau between even-odd(EO) state and the chiral state is something not 
detected in the conventional DMRG calculation up to 382 sites\cite{okunishi03,okunishi08}. 
Such anomaly is indeed blurred in the usual numerical calculation due to discreteness or finite size effect, 
particularly when it is related to the ordered state with incommensurate wave number, 
e.g., the present chiral state which arose due to strong geometrical frustration. 
\par
%
\begin{figure*}[tbp]
\begin{center}
\includegraphics[width=17.5cm]{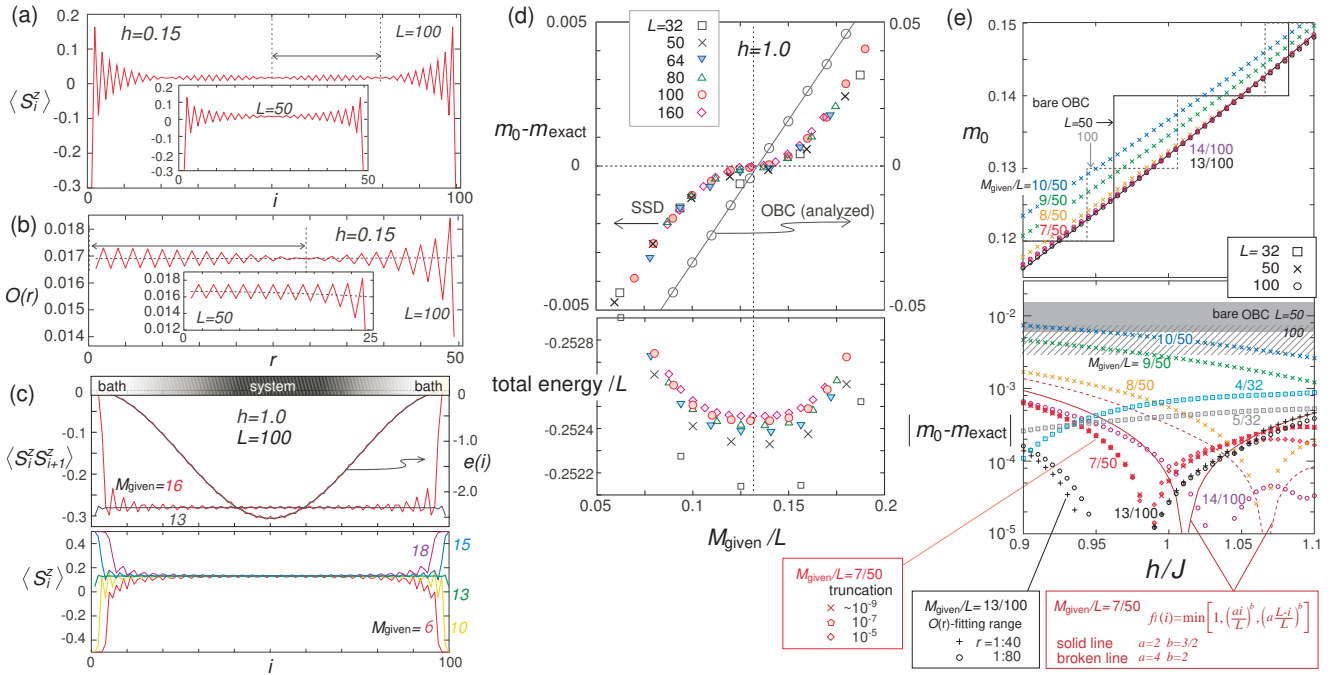}
\end{center}
\caption{Analyses given for the Heisenberg spin chain. 
(a) Site ($i$)-dependent local spin operator, $\langle S^z_i\rangle$, 
and (b) the average of $\langle S^z_i\rangle$ over $2r$-sites from the system center, $O(r)$ (Eq.(\protect\ref{mr})) at $h/J=0.15$ for $L=50$ and $100$. 
Broken lines in (b) are the fitting function $O(r)=m_0+c r^2$. 
(c) Site($i$)-dependence of $\langle S^z_iS^z_{i+1}\rangle$, and 
deformed bond energy $e(i)=3Jf_0(i) \langle S^z_iS^z_{i+1}\rangle$ at $h/J=1$. 
Lower panel shows $\langle S^z_i\rangle$ for several choices of $M_{\rm given}$. Excess/deficient $\langle S^z_i\rangle$ is absorbed at the edge "bath". 
(d) $m_0-m_{\rm exact}$ ($m_{\rm exact}$: exact solution) and the corresponding total energy(lower panel)
as functions of $M_{\rm given}/L$ for various $L$. Those analyzed under OBC is compared. 
(e) $m_0$ and the logarithmic $|m_0-m_{\rm exact}|$ as a function of $h$ for various $M_{\rm given}/L$. 
Shaded ($L=50$) and meshed ($L=100$) region are the typical accuracies of the bare OBC results (stepwise data on the upper panel). 
Data for different truncation errors, different fitting range ($r$), and with power scaling functions (non-SSD $f_l(i)$) are shown together. 
}
\label{f2}
\end{figure*} 
The analyses to obtain the above results are extremely simple. 
We fix the value of $h$, give appropriate but not necessarily precise total number of the $z$-component of spins, $M_{\rm given}$, 
and perform a usual DMRG on ${\mathcal H}_{\rm deform}$(here we take ${\mathcal H}_{\rm SSD}$) to calculate the expectation value of the local operator, $\langle S^z_i \rangle$. 
The representative behavior of the spacial dependence of $\langle S^z_i \rangle$ is shown in Fig.~\ref{f2}(a) for the Heisenberg chain under magnetic field. 
There arises two types of spacial oscillations with short and long periods with a particularly large peak at the edge. 
The center line of the oscillation near $i \sim L/2$ is almost flat, which is expected to {\it reproduce the value of $\langle S^z_i \rangle$ 
realized in the bulk limit}, $\langle S^z\rangle_{L=\infty}$. 
One could extract systematically the value of $\langle S^z\rangle_{L=\infty} $ as in the following. 
We first take the average of $\langle S^z_i \rangle$ from the center of the system toward the edge over $2r$ sites as, 
\begin{equation}
O(r)= \frac{1}{2r}\sum_{i=-r}^{r-1} \big\langle S^z_{i+\frac{L}{2}}\big\rangle, 
\label{mr}
\end{equation}
as shown in Fig.~\ref{f2}(b). Reflecting the non-uniform structure of $\langle S^z_i\rangle$ throughout the system, 
$O(r)$ oscillates from $r=1$, with coexisting large and small periods. 
One could safely fit $O(r)$ by setting the fitting range from one node of the large oscillation to another. 
The broken line in Fig.~\ref{f2}(b) is the result fitted in powers of $r$ as, $O(r)=m_0 + c r^2$\cite{howtofit}. 
The resultant $m_0$ give the magnetization density, $M/L$, 
shown in Fig.\ref{f1}, in almost perfect coincidence with the exact solution even for system size as small as $L=32$, 
a relatively small size which could be calculated even by the exact diagonalization. 
The obtained results are thus arbitrary continuous real numbers, and are free from finite size discreteness. 
\par
Let us discuss the implication of the success in taking $m_0\!=\!\langle S^z\rangle_{L=\infty}$. 
Figure~\ref{f2}(c) shows the expectation value of the nearest neighbor spin-spin interaction, $\langle S^z_i S^z_{i+1}\rangle$. 
The results for $M_{\rm given}\!=\!13$ is almost uniform, whereas for $M_{\rm given}\!=\!16$ a large oscillation amplitude is found near the edges. 
However, the bond energy, $e(i)= J f_1(i)\langle S_i \cdot S_{i+1}\rangle$, which is scaled down smoothly from the center toward the edge, 
are almost equivalently smooth for both cases (see Fig.~\ref{f2}(c)). 
This is because the energy of extra spins, $\Delta M\!=\!M_{\rm given}\!-\!m_0L$ 
($\sim 3$ for $M_{\rm given}\!=\!16$), which concentrate near the edge sites, remains almost zero, 
so that they can be approximately excluded from the main part of the system. 
Thus, one could effectively get rid of the excess $\Delta M>0$ by concentrating them at the edge sites, 
or supply the deficient $\Delta M<0$ from the edge site to the center of the system (see the lower panel of Fig.~\ref{f2}(c)).
The variational determination of a wave function in standard DMRG automatically carries out this procedure. 
In other words, the edge site serves as a small particle bath which is connected smoothly with the main part of the system, 
and the distribution of the spins/particles are automatically optimized by the variational principle. 
One may regard this as a "grand canonical" setup, 
in analogy with the grand canonical ensemble used for convenience in quantum mechanical problems 
when it is not easily solved by fixing the number of particles in the system for technical reason. 
Here, the finite $L$ requires smooth variation of non-integer particle number under applied field, 
which cannot be given by hand as a conserved number. 
Instead, by loosely dividing the system into energetically inequivalent center and the edges by the scaling function $f_l(i)$, 
and allowing for a small quantum fluctuation of energy and particle numbers between them via $-f_1(j)h_1(j)$ ($j$:near the edges),  
one could obtain a non-integer expectation number as a quantum ensemble. 
The range of the ensemble required to adjust the particle density within the order of $1/L$ near the system center, 
does not need to be large. Namely, a small bath connected with the small fluctuation is enough. 
\par
One thus expects that $m_0$ does not depend on the choice of $M_{\rm given}$ owing to the buffer edges. 
Figure~\ref{f2}(d) shows $m_0$ at fixed $h$ as a function of $M_{\rm given}$. 
There exists an inflection point in the very vicinity of the exact solution $m_{\rm exact}$, 
which remains almost unchanged with $L$. 
This point, giving with most precise result, can be detected practically as $M_{\rm given}$ having the minimum of the total energy. 
This fact also verifies the variational principle we discussed earlier.  
We also perform the same analysis under the OBC {\it without deformation}. 
Notice that the scale of the vertical axis is ten times larger than the deformed results, 
and that $m_0$ is a linearly increasing function of $M_{\rm given}/L$, i.e. with no inflection point. 
This comparison guarantees that not the translational symmetry breaking itself, but the spatial scaling down of the Hamiltonian, 
is important to endow the edge site a role of particle bath. 
\par
We further examined the accuracy of the evaluated $m_0$ as a function of $h$ for various $M_{\rm given}/L$, as shown in Fig.~\ref{f2}(e). 
Regardless of the value of $M_{\rm given}$, $m_0$ is a smooth function of $h$, 
and the deviation from the exact results remains less then $\sim O(10^{-4})$ over a relatively wide range of $h \sim 0.1$
when $|M_{\rm given}/L-m_{\rm exact}|\lesssim 0.1$. 
Other factors do not deteriorate the results as well(see the lower panel of Fig.~\ref{f2}(e)); 
the degree of accuracy also do not depend much on the fitting range of Eq.(\ref{mr}), truncation errors of the DMRG, 
or on the details of $f_l(x)$ (non-SSD functions). 
Indeed, the magnetization curve in Fig.~\ref{f1} is obtained without optimizing $M_{\rm given}$ to the energy minimum or inflection point; 
one could obtain a reasonably accurate result by just performing a preliminary set of calculation to determine approximately proper $M_{\rm given}$ in advance. 
\par
In the 1D quantum spin systems, a method called the product wave function renormalization group\cite{okunishi97} 
reached the relatively smooth magnetization curve consistent with Fig.~\ref{f2}(b). 
This method shares common concept with the so-called matrix product states (MPS), 
which is recently applied to the imaginary time-evolving block decimation, 
a method having translationally invariant wave function by construction\cite{vidal07}. 
However, the MPS description is so far realistically applied to particular quantum spin systems in 1D or 2D, 
and further, the target states are no longer well described by the MPS in the vicinity of the quantum critical point. 
\par
%
%
\begin{figure}[tbp]
\begin{center}
\includegraphics[width=7.5cm]{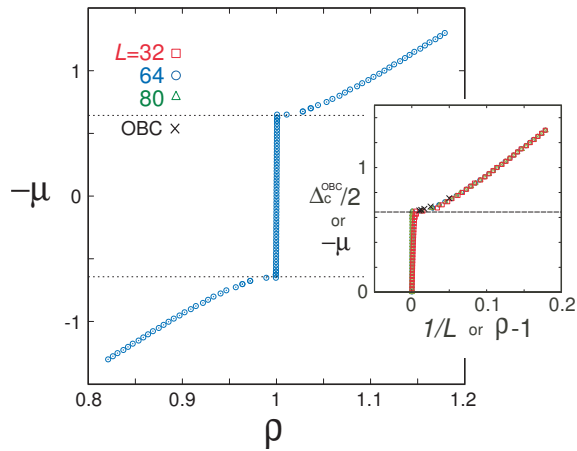}
\end{center}
\caption{(a) Chemical potential $\mu$ as a function of electron density, $\rho$, 
by the present analyses with DMRG on SSD Hubbard Hamiltonian with $m\le 300$ at $L=64$ for $U/t=4$. 
Broken line indicates the one particle gap of the exact solution\cite{exact}. 
Inset shows the evaluation of one particle gap by the standard finite size scaling procedure using the DMRG results for the usual OBC without deformation, 
in comparison with the present SSD results with $L=32, 64, 80$. 
}
\label{f3}
\end{figure} 
\par
By contrast, the present analyses is applied to the strongly correlated electronic system as well. 
We choose the one-dimensional Hubbard model as a typical example, 
with $u(i)=U (n_{i\uparrow}\!-\!1/2)(n_{i\downarrow}\!-\!1/2)$ and 
$h_1(i)=-t \sum_{\sigma=\uparrow,\downarrow}(c_{i\sigma}^\dagger c_{i+1\sigma}+{\rm H.c.})$, 
where $c_{i\sigma}^\dagger/c_{i\sigma}$ are the creation/annihilation operator of electron on site-$i$ with spin-$\sigma$. 
At half-filling, the model exhibits a Mott insulating state for any finite value of $U/t$, 
where $U$ and $t$ denote the on-site Coulomb interaction and transfer integral, respectively. 
This time, the chemical potential, $-\mu$, is varied under the appropriately given total electron number, $N_e^{\rm given}$, 
and fitting the mean value of electron density around the center of the system as, 
\begin{equation}
O(r)= \frac{1}{2r}\sum_{i=-r}^{r-1} \langle n_{i\uparrow}+n_{i\downarrow} \rangle = \rho +c r^2, 
\end{equation}
will give the electron density in the bulk limit, $\rho$, for any value of $\mu$. 
Figure~\ref{f3} shows the $\mu$-$\rho$ curve at $U/t=4$ for $L=64$, which clearly indicates kinks at $\mu \sim \pm \Delta_c (L=\infty)/2$, 
where $\Delta_c (L=\infty) = 1.2867(2)$ is the charge gap evaluated from the exact solution\cite{exact}. 
Again, the curve is continuous and the typical finite size effect remains less than the order of 10$^{-4}$
(see the comparison of $L=32,64$ and 80 in the inset). 
We also calculated the charge gap, $\Delta_c^{\rm (OBC)}(L)$, under usual OBC without deformation, and plotted them together in the inset of Fig.\ref{f3} 
by sharing the two axes, $\rho$ and $-\mu$, with $1+L^{-1}$ and $\Delta_c^{\rm (OBC)}(L)/2$, respectively, 
due to the following context; 
the states at $|\mu| \ge \Delta_c/2$ off the gap are the Tomonaga-Luttinger liquid, where the low energy excitation is dominated by 
the bosonic quasi-particles which can be approximated by the non-interacting fermions in a 1D chain in the dilute limit. 
In calculating $\Delta_c(L)$ with electron number, $N_e^{\rm given}=L+1$, 
the density of the doped quasi-particle corresponds to $1/L$, and $\Delta_c(L)$ as a function of $1/L$ will approximately give 
its low energy dispersion, namely the bulk $\mu$-$\rho$ curve. 
Indeed, as shown in the inset of Fig.~\ref{f3}, $\Delta_c(L)$ as a function of $1/L$ show good correspondence with the $\mu$-$\rho$ curve by our analyses, 
which means that our grand canonical analysis on small systems well reproduce the bulk properties.
\par
To summarize, we elucidated a way to directly obtain bulk physical quantities against continuously varying field in a small finite size cluster. 
We find that scaling down the energy of the Hamiltonian from the system center toward the edges endows to the edge state a role as a small particle bath. 
The particles trapped at the edges have negligibly small energy, and are connected to the bulk part by an ideally small fluctuation. 
The variational optimization of the wave function actually uses the edge sites as particle bath, 
and one obtains continuously varying conserved numbers $S^z$ or $N_e$ in the main part of the system. 
The results obtained are almost free from finite size effect and reproduces the exact solution within the accuracy of $\sim 10^{-4}$ 
even for the system as small as $L \sim {\mathcal O}(10)$. 
The present analyses is applied to methods other than DMRG, such as exact diagonalization, quantum Monte Carlo method, 
or other variational methods, as far as the optimized wave function is used, 
and thus will open a new path toward solving numerous still unknown problems in low-energy many body physics.



\begin{references}
\bibitem{howtofit} The fitting becomes more accurate by taking wider range. 
Since the period of short oscillation may influences the results, 
better operation is given by taking nodes of the large oscillation as a fitting period, which may not necessarily include $r\sim 1$. 
The term $c r^2$ is required to stabilize the fitting, but its value is small, ${\mathcal O}(10^{-5})\sim {\mathcal O}(10^{-7})$.
\end{references}
\end{document}